\documentclass[a4paper]{article}

\usepackage{INTERSPEECH2022}
\usepackage{algorithm}
\usepackage{algpseudocode}
\usepackage{float}
\usepackage{graphicx}

\title{Acoustic-to-articulatory Speech Inversion with Multi-task Learning
}
\name{Yashish M. Siriwardena$^1$, Ganesh Sivaraman$^2$, Carol Espy-Wilson$^1$}
\address{
  $^1$University of Maryland College park, MD, USA\\
  $^2$Pindrop, GA, USA}
\email{yashish@terpmail.umd.edu, ganesa90@gmail.com, espy@umd.edu}

\begin{document}

\maketitle
\begin{abstract}
Multi-task learning (MTL) frameworks have proven to be effective in diverse speech related tasks like automatic speech recognition (ASR) and speech emotion recognition. This paper proposes a MTL framework to perform acoustic-to-articulatory speech inversion by simultaneously learning an acoustic to phoneme mapping as a shared task. We use the Haskins Production Rate Comparison (HPRC) database which has both the electromagnetic articulography (EMA) data and the corresponding phonetic transcriptions. Performance of the system was measured by computing the correlation between estimated and actual tract variables (TVs) from the acoustic to articulatory speech inversion task. The proposed MTL based Bidirectional Gated Recurrent Neural Network (RNN) model learns to map the input acoustic features to nine TVs while outperforming the baseline model trained to perform only acoustic to articulatory inversion.  

\end{abstract}
\noindent\textbf{Index Terms}: acoustic-to-articulatory speech inversion, multi-task learning, acoustic-to-phoneme mapping, biGRNNs

\section{Introduction}

Human speech production is a highly complex task which involves synchronized motor control of speech articulators.   The inverse problem of determining the trajectories of the movement of speech articulators from the speech signal is referred to as acoustic-to-articulatory speech inversion \cite{Sivaraman_ASA, SI_Trans_ppr}.  This mapping from acoustics to articulation is an ill-posed problem which is known to be highly non-linear and non-unique \cite{Qin2007}. However, developing Speech Inversion (SI) systems have gained attention over the recent years mainly due to its potential in a wide range of speech applications like Automatic Speech Recognition (ASR) \cite{Frankel2001ASRA, Mitra2010, Mitra2011}, speech synthesis \cite{speech_synthesis_1, speech_synthesis_2}, speech therapy \cite{Fagel2008A3V} and most recently with detecting mental health disorders like Major Depressive Disorder and Schizophrenia \cite{espywilson19_interspeech,Siriwardena_SZ}. Real articulatory data are collected by techniques like X-ray microbeam \cite{Westbury1994a}, Electromagnetic Articulometry (EMA) \cite{Schonle1987} and real-time Magnetic Resonance Imaging (rt-MRI) \cite{Narayanan2004}. All these techniques are expensive, time consuming and need specialized equipment for observing articulatory movements directly \cite{Sivaraman_ASA}. This explains why developing a speaker-independent SI system that can accurately estimate articulatory features for any unseen speaker is of greater need. 

Over the past few years, deep neural network (DNN) based models have propelled the development of SI systems to new heights. Bidirectional LSTMs (BiLSTMS) \cite{illa18_interspeech, illa20_interspeech}, CNN-BiLSTMs \cite{Shahrebabaki2020, illa_CNN_BLSTM}, Temporal Convolutional Networks (TCN) \cite{shahrebabaki21_interspeech} and transformer models \cite{udupa21_interspeech} have gained state-of-the-art results with multiple articulatory datasets \cite{Tiede2017}. To further improve the speech inversion task, people have tried incorporating phonetic transcriptions as an input along with acoustic features \cite{Singh2020ACS, Shahrebabaki2020}. One of the limitations of these models is that you need phonetic transcriptions of the speech audio file at the time of inference. To address this issue while also leveraging on the additional information that phonetic transcriptions offer, we propose a Bidirectional Gated Recurrent Neural Network (BiGRNN) model, implemented with a multi-task learning framework, to perform acoustic-to-articulatory speech inversion. The MTL based model does not need phonetic transcriptions at the time of inference, but benefits from learning the mapping from acoustics-to-phonetics to improve generalizability of SI systems. 

The key contributions of the paper can be listed as follows :  
\begin{itemize}
  \item We propose a MTL based BiGRNN model to perform acoustic-to articulatory speech inversion by also learning a shared task of acoustic-to-phoneme inversion.
  \item We compare and contrast two training algorithms to optimize the proposed MTL model
  \item By conducting an ablation study we assert the importance of doing multi-task learning for speech inversion
\end{itemize}
% we propose a MTL based BiGRNN model to perform acoustic-to articulatory speech inversion by also learning a shared task of articulatory-to-phoneme inversion. We implement two training algorithms to optimize the MTL model and do an ablation study to assert the importance of doing multi-task learning for speech inversion.  

\section{Speech Inversion System}

\subsection{Multi-task Learning : Related Work}

The idea of Multi-task learning (MTL) was formally presented by Caruana et al. \cite{Caruana_MTL} as an inductive transfer mechanism with the principle goal of improving generalization capability of Machine Learning (ML) models. MTL helps improve generalizability of ML models by leveraging domain-specific information of training data which can be used in related tasks. Effectively, what happens is that the training data for the parallel task serve as an inductive bias \cite{Caruana_MTL}. MTL has also been utilized as a solution for the data sparsity problem where one task has a limited number of labeled data and training individual models for each task is difficult. From this perspective, MTL is a useful tool which can reuse the existing knowledge and reduce the cost of collecting challenging datasets (e.g. articulatory datasets). The secret behind the success of MTL lies with the use of more data from different learning tasks compared to learning a single task, hence learning better representations and reducing the risk for overfitting \cite{MTL_survey}. 

MTL has widely been used in computer vision and a recent work \cite{MTL_vision} has implemented a MTL model to work on 12 different datasets while achieving the state-of-the-art with 11 of them. MTL has also been explored in Automatic Speech Recognition (ASR) tasks \cite{Kim2017JointCB, Hori2017AdvancesIJ}, text-to-speech (TTS) \cite{fastspeech2} and in speech emotion recognition (SER) \cite{SER_MTL1, SER_MTL2}. Cai et al \cite{SER_MTL2} recently presented the state-of-the-art results for the SER task with IEMOCAP dataset using their model based on a MTL framework. 

\subsection{Dataset Description}

We used the Haskins Production Rate Comparison (HPRC) database which contains recordings from 4 female and 4 male subjects reciting 720 phonetically balanced IEEE sentences \cite{IEEE_sentences} at normal and fast production rates \cite{Tiede2017}. The recordings were done using a 5-D electromagnetic articulometry (EMA) system (WAVE; Northern Digital). First, every sentence was produced at speaker’s preferred ‘normal’ speaking rate and then a ‘fast’ repetition of the same, without making errors. Sensors were placed on the tongue (tip (TT), body (TB), root (TR)), lips (upper (UL) and lower (LL)) and mandible, together with reference sensors on the left and right
mastoids, and upper and lower incisors (UI, LI). These EMA trajectories were obtained at 100 Hz and then were low-pass filtered at 5 Hz for references and 20 Hz for articulator sensors.
Synchronized audio was recorded at 22050 Hz. The following geometric transformations were used to obtain 9 TVs (namely  Lip Aperture (LA), Lip Protrusion (LP), Tongue Body Constriction Location (TBCL), Tongue Body Constriction Degree (TBCD), Tongue Tip Constriction Location (TTCL), Tongue Tip Constriction Degree (TTCD), Jaw Angle (JA), Tongue Middle Constriction Location (TMCL) and Tongue Middle Constriction Degree (TMCD)). The equations to compute the geometric transformations are presented in our previous work in \cite{Sivaraman_ASA, seneviratne19_multicorpus}.  

\begin{figure}[t]
  \centering
  \includegraphics[width=1.0\columnwidth]{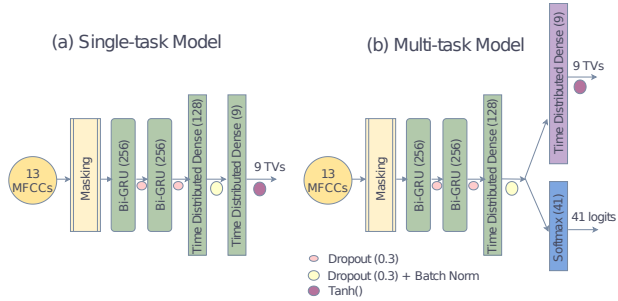}
  \caption{Single-task and Multi-task model architectures}
    \label{fig:model_archi}
\end{figure}

\subsection{Audio Features}

All the audio files from the HPRC dataset (both `normal' and `fast' rate) are first segmented into 2 second long segments and the shorter audios are zero padded at the end. Previous studies with developing SI systems have shown MFCCs to be superior over conventional Melspectrograms and Perceptual Linear Predictions (PLPs) as acoustic features \cite{Sivaraman_ASA}. Based on that, in this study, we use Mel-Frequency Cepstral Coefficients (MFCCs) as the input audio feature for the proposed SI systems. MFCCs are extracted using a 20ms Hamming analysis window with a 10ms frame shift. 13 cepstral coefficients were extracted for each frame while 40 Mel frequencies were used. Each MFCC was utterance wise normalized (z-normalized) prior to model training. 

\subsection{Phoneme Features}

The HPRC dataset contains phonetic alignment for the recorded utterances. The phone alignment is extracted using the Penn Phonetics Lab Forced Aligner(P2FA)\footnote[1]{https://github.com/jaekookang/}. We remove the allophonic variations of the monophones and retain only 40 monophone units.
Using the forced alignment we created frame wise monophone labels for all of the HPRC dataset. The one-hot encoded frame-wise monophone labels are the phonetic features used in this study. 

% \footnote[1]{https://github.com/jaekookang/p2fa_py3}

\subsection{Model Architecture}

In this paper, we use a novel Bidirectional Gated Recurrent Neural Network (BiGRNN) model to implement both the single-task and multi-task SI systems. Both the single-task and multi-task models have the same backbone which includes 3 bidirectional layers of Gated Recurrent Units (GRUs) followed by a time distributed fully connected layer. Single-task model which predicts TVs has an additional time distributed fully connected layer to predict the TVs (output layer). On the other hand, the multi-task model has two output layers, one a time distributed fully connected layer to predict the TVs and the other a softmax layer to predict phoneme labels. Figure \ref{fig:model_archi} shows the architecture of the single task model on the left and the multi-task model on the right. 

\subsection{Performance Measurements}

All the models are evaluated with the Pearson Product Moment Correlation (PPMC) scores computed between the estimated TVs and the corresponding ground-truth TVs. Equation \ref{eq1} is used to computed the PPMC score, where $X$ represents the estimated TVs, $\overline{X}$ the mean of the estimated TVs, $Y$ the ground-truth TVs, $\overline{Y}$ the mean of the ground-truth TVs and $N$ the number of TVs.
\begin{equation}
  PPMC = \frac{\sum_i^N{(X[i] - \overline{X})(Y[i] - \overline{Y})}}{\sqrt{\sum_i^N{(X[i] - \overline{X})^2(Y[i] - \overline{Y})^2}}}
  \label{eq1}
\end{equation}

\begin{table*}[t]
    \centering
    \normalsize
     \caption{Single-task vs Multi-task learning for TV predictions}
    % \resizebox{\columnwidth}{!}
    \begin{tabular}{r r r r r r r r r r r}
    \toprule
    \multicolumn{1}{c}{\textbf{Model}} & 
                                    \multicolumn{1}{c}{\textbf{LA}}  &      \multicolumn{1}{c}{\textbf{LP}}  &  \multicolumn{1}{c}{\textbf{JA}}&  \multicolumn{1}{c}{\textbf{TTCL}}&  \multicolumn{1}{c}{\textbf{TTCD}}&  \multicolumn{1}{c}{\textbf{TMCL}}&  \multicolumn{1}{c}{\textbf{TMCD}}& 
                                    \multicolumn{1}{c}{\textbf{TBCL}}& 
                                    \multicolumn{1}{c}{\textbf{TBCD}}& \multicolumn{1}{c}{\textbf{Average}} \\
    \midrule
    Single-task & 0.764 & 0.661 & 0.790 & 0.706 & 0.778 & 0.741 & 0.801 & 0.725 & 0.742 & 0.745\\
    Multi-task (Algo 1) & 0.792 & 0.681 & 0.796 &0.747 &0.793 &0.775 &0.799 &0.760 &0.764 &0.767\\
    Multi-task (Algo 2) & 0.794 &0.680 &0.806 &0.741 &0.797 &0.775 &0.806 &0.762 &0.766 & \textbf{0.770}
    \\\bottomrule
    \end{tabular}
    \label{table: single_vs_multitask}
\end{table*}

\vspace*{-5pt}
\subsection{Model training}

The HPRC dataset was divided into training, development, and testing sets, so that the training set has utterances from 6 speakers (3 Males, 3 Females) and the development and testing sets have utterances of 2 speakers (1 male,1 female) equally split between them. None of the subjects in training are present in the development and testing sets and hence all the models are trained in a `speaker-independent' fashion. The split also ensured that around 80\% of the total number of utterances were present in the training, and the development and testing sets have a nearly equal number of utterances. This allocation was done in a completely random manner.  

All the models were implemented with Tensorflow-Keras machine learning framework and trained with NVIDIA TITAN X GPUs. For all single-task and multi-task models, ADAM optimizer with a starting learning rate of 1e-3 and an exponential learning rate scheduler was used. The starting learning rate was maintained up to 10 epochs and then decayed exponentially after each subsequent 5 epochs. To choose the best starting `learning rate' (LR), we did a grid search on [1e-3, 3e-4, 1e-4] and to choose the training batch size we did a similar grid search on [16,32,64,128]. The best PPMC scores were obtained for 1e-3 as the LR and 128 as the batch size for training.

\subsection{Training Paradigms for multi-task SI systems}

We experimented with two distinct training algorithms to optimize the MTL model. We denote the input MFCC features to the model as $x\epsilon R^{L\times d}$ where $L$ (=200) is the number of samples in each utterance and $d$ (=13) is the number of MFCCs. Let $f_{\phi}$ be the mapping from MFCCs to TVs from the multi-task model where $\phi$ defines the shared model parameters to be learned. Similarly, let $g_{\phi}$ be the mapping from MFCCs to phoneme logits. Then the output TV prediction from the TV output layer $\hat{y}_{tv} \epsilon R^{L \times T}$ can be defined from equation \ref{eq2} and similarly the output logits from the phoneme prediction, $\hat{y}_{ph} \epsilon R^{L \times V}$ can be defined from eqaution \ref{eq3}. Here $T$ (=9) is the number of TVs predicted and $V$ (=41) is the number of phonemes in the dictionary + the symbol for zeros (padded for shorter utterances). We used the Mean Absolute Error (MAE) loss between ground truth TVs $y_{tv}$ and predicted TVs $\hat{y}_{tv}$ and cross entropy error loss between ground truth one-hot encoding labels of phonemes $y_{ph}$ and the predicted phonemes $\hat{y}_{ph}$.

\begin{equation}
  \hat{y}_{tv} = f_{\phi}(x) \: ; \: x\epsilon R^{L\times d}
  \label{eq2}
\end{equation}

\begin{equation}
  \hat{y}_{ph} = g_{\phi}(x) \: ; \: x\epsilon R^{L\times d}
  \label{eq3}
\end{equation}

\vspace*{-6pt}
\subsubsection{Training Algorithm 1}

Here the multi-task model is optimized for each task in an alternating fashion. In each epoch, the model weights $\phi$ are first learned from the TV prediction task and the learned weights are then used for computing phoneme labels $\hat{y}_{ph}$. The final model weights $\phi[i]^{*}$ are then updated with the phoneme prediction task and the process is repeated for the given number of $Epochs$. 

\begin{algorithm}
\caption{Iterative Loss Optimization}\label{alg:tr2}
\begin{algorithmic}
\Require: $x \epsilon R^{L\times d}, y_{ph}, y_{tv}, Epochs (\epsilon R)$
% \State $Input = x$
% \State $X \gets x$
% \State $N \gets n$
\While{$i < Epochs $}
    \State $ \hat{y}_{tv} \gets f_{\phi[i-1]}(x)$
    \State $ L_{tv} \gets MAE(\hat{y}_{tv}, y_{tv})$
    \State $ \phi[i] \gets \min_{\phi} L_{tv}$
    \State $ \hat{y}_{ph} \gets g_{\phi[i]}(x)$
    \State $ L_{ph} \gets CrossEntropy(\hat{y}_{ph}, y_{ph})$
    \State $ \phi[i]^{*} \gets \min_{\phi} L_{ph}$

    \State $ i \gets i+1 $

% \EndIf
\EndWhile
\end{algorithmic}
\end{algorithm}

\vspace*{-6pt}
\subsubsection{Training Algorithm 2}

In this training algorithm we optimize a joint loss $L_{joint}$, where the phoneme prediction loss $L_{ph}$ is weighted to combine with the TV prediction loss $L_{tv}$. The contribution of $L_{ph}$ is controlled by the weight $\alpha \epsilon (0,1)$, which is a hyper-parameter to be tuned. Here the model is trained with an early stopping criteria monitoring the validation loss ($ValLoss$) with a patience $p$ (=10). 

\begin{algorithm}
\caption{Joint Loss Optimization}\label{alg:tr1}
\begin{algorithmic}
\Require: $x \epsilon R^{L\times d}, ValLoss, p \epsilon R, \alpha (0<\alpha<1), y_{ph}, y_{tv}$
% \State $Input = x$
% \State $X \gets x$
% \State $N \gets n$
\While{$Val Loss[i]< Val Loss [i-p]$}
% \If{$N$ is even}
    \State $ \hat{y}_{ph} \gets g_{\phi[i-1]}(x)$
    \State $ \hat{y}_{tv} \gets f_{\phi[i-1]}(x)$
    \State $ L_{ph} \gets CrossEntropy(\hat{y}_{ph}, y_{ph})$
    \State $ L_{tv} \gets MAE(\hat{y}_{tv}, y_{tv})$
    \State $ L_{joint} \gets L_{tv} + \alpha L_{ph}$
    \State $ \phi[i] \gets \min_{\phi} L_{joint}$
    \State $ i \gets i+1 $

% \EndIf
\EndWhile
\end{algorithmic}
\end{algorithm}

\vspace*{-6pt}
\section{Experiments and Results}

\subsection{Single-task vs Multi-task Learning for TV prediction}

\begin{figure}[t]
  \centering
  \includegraphics[width=1.0\columnwidth, height=70mm]{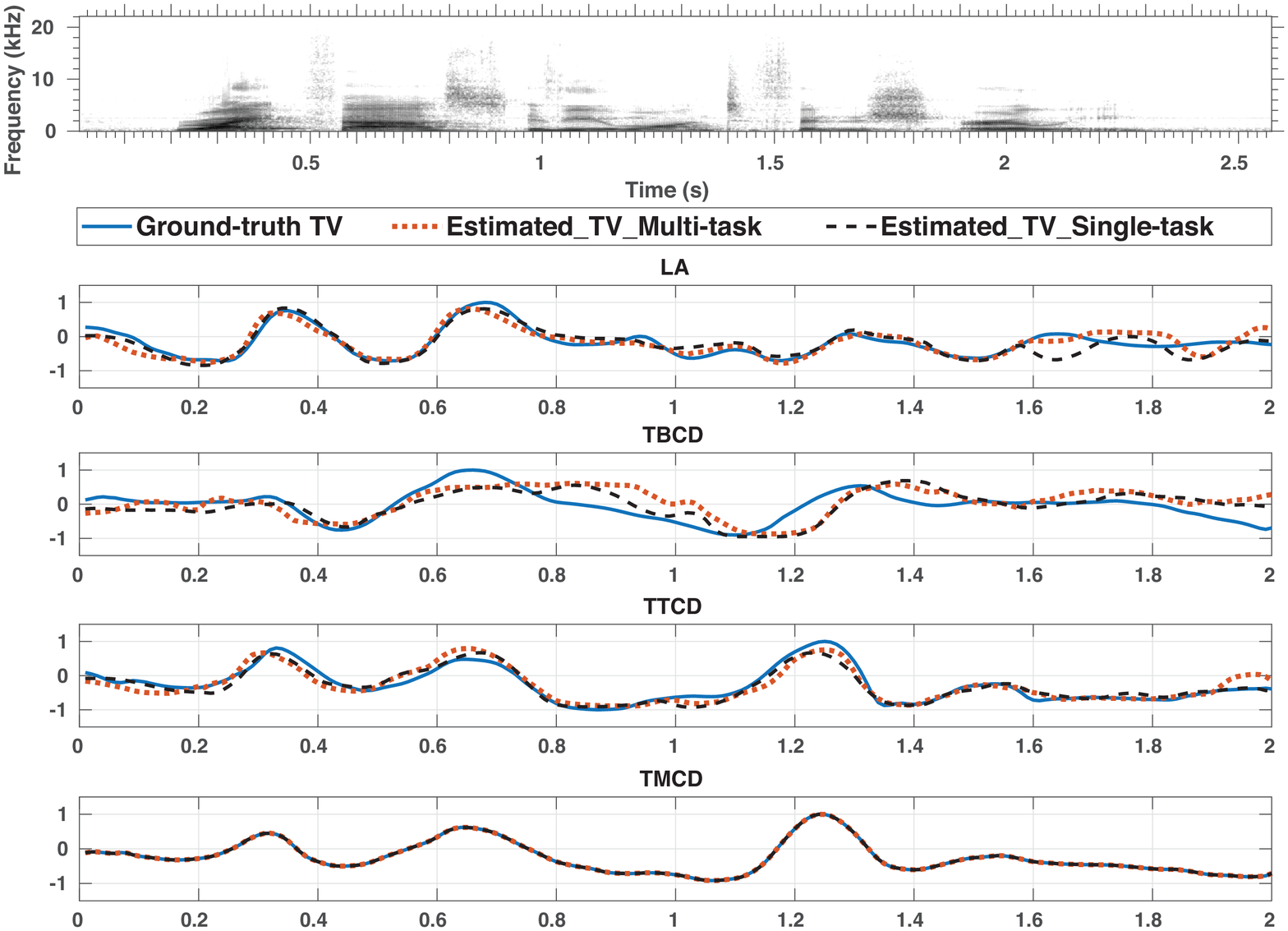}
  \caption{LA and constriction degree TV plots for the utterance `Write fast if you want to finish early' estimated using Multi-task model and the Single-task model.
Solid blue Line - actual TV (from HPRC database), red dotted line - estimated TV from Multi-task model, black dashed Line
- estimated TV from Single-task model}
    \label{fig:tv_plots}
\end{figure}

\begin{figure}[t]
  \centering
  \includegraphics[width=1.0\columnwidth, height=70mm]{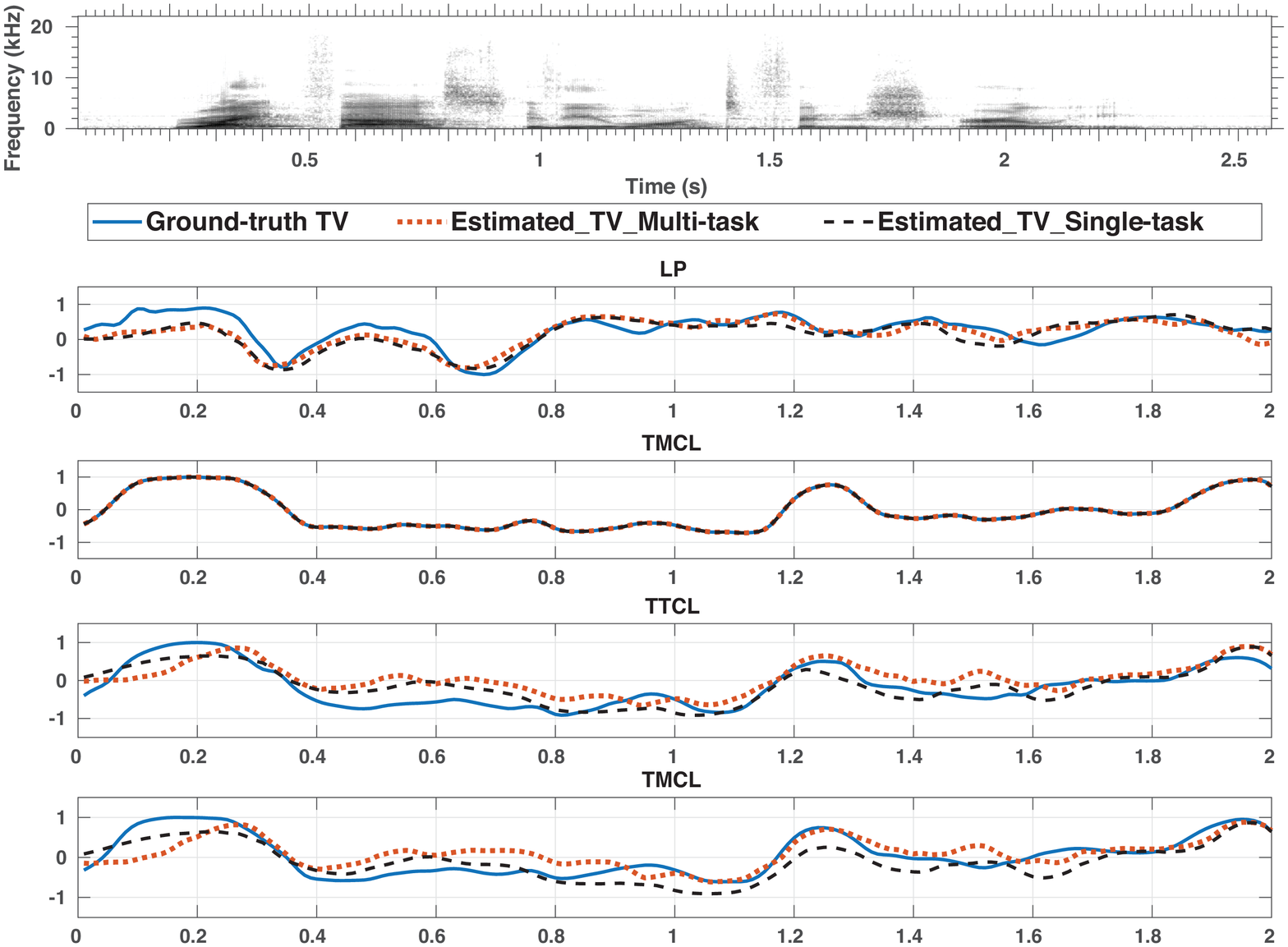}
  \caption{LP and constriction location TV plots for the utterance `Write fast if you want to finish early' estimated using Multi-task model and the Single-task model.
Solid blue Line - actual TV (from HPRC database), red dotted line - estimated TV from Multi-task model, black dashed Line
- estimated TV from Single-task model}
    \label{fig:tv_plots_2}
\end{figure}

Table \ref{table: single_vs_multitask} shows the results of the single-task model when compared to the two multi-task models trained with two training algorithms. The reported PPMC scores are from evaluations of the speaker-independent test set. Figure \ref{fig:tv_plots} shows ground-truth TVs and predicted TVs, LA, TBCD, TTCD and TMCD for an example utterance estimated by the multi-task and the single-task models. Figure \ref{fig:tv_plots_2} shows ground-truth TVs and predicted TVs, LP, TBCL, TTCL and TMCL for the same utterance estimated by the multi-task and the single-task models.  
 
\vspace*{-6pt}
\subsection{Baseline comparison on previous work with HPRC dataset}

Table \ref{table: baseline_models} lists the reported PPMC scores of the work in Shahrebabaki et. al. \cite{SI_Trans_ppr, Shahrebabaki2020} for speaker-independent SI task using the HPRC dataset. Both the studies and our study use same target TVs (derived from same transformations \cite{Sivaraman_ASA}) and use both `normal' and `fast' rate utterances without any speaker rate matching at evaluations. The comparison is still not a perfect one given that there can be differences in the test splits used for evaluation.       

\vspace*{-6pt}
\begin{table}[H]
  \caption{Baseline models with HPRC dataset}
  \centering
  \resizebox{\columnwidth}{!}{
  \begin{tabular}{l c}
    \toprule
    \multicolumn{1}{c}{\textbf{Model}} & \multicolumn{1}{c}{\textbf{Average PPMC score}} \\
    \midrule
    Feed-forward model$^{*}$ \cite{SI_Trans_ppr} & 0.705 \\
    CNN-BiLSTM model$^{*}$ \cite{Shahrebabaki2020} & 0.755\\
    Single-task BiGRNN model & 0.745 \\
    Proposed Multi-task BiGRNN model & 0.770
    \\\bottomrule
  \end{tabular}
  \label{table: baseline_models}}
\end{table}

\vspace*{-10pt}
\subsection{Ablation Study}
\label{sec:ablation}

We changed the weight $\alpha$ in the MTL model trained with algorithm 2 to explore how the phoneme learning task would help the desired SI task. Recall that $\alpha$ controls the amount of contribution from the phoneme prediction loss $L_{ph}$ to the joint loss $L_{joint}$. Here setting $\alpha = 0$ is equivalent to the single-task model.

\vspace*{-10pt}
\begin{table}[H]
  \caption{Contribution of phoneme learning task for the SI task}
  \centering
  \resizebox{\columnwidth}{!}{\begin{tabular}{r c c}
    \toprule
    \multicolumn{1}{c}{} & \multicolumn{1}{c}{\textbf{Average PPMC}}  &   \multicolumn{1}{c}{\textbf{Phoneme Accuracy (\%)}} \\
    \midrule
    $\alpha =0.0$ & 0.743 & 2.25   \\
    $\alpha =0.1$ & 0.762 & 70.60 \\
    $\alpha =0.3$ & 0.766 & 72.53 \\
    \boldsymbol{$\alpha =0.5$} & \textbf{0.770} & 72.88 \\
    $\alpha =0.8$ & 0.759 & 72.90 \\
    $\alpha =1.0$ & 0.758 & 73.60
    \\\bottomrule
  \end{tabular}}
  \label{table: ablation_tb1}
\end{table}

\vspace*{-10pt}
\begin{table}[H]
  \caption{Training Time : Single-task and Multi-task models}
  \centering
  \resizebox{\columnwidth}{!}
  {\begin{tabular}{l c c}
    \toprule
    \multicolumn{1}{c}{\textbf{Model Type}} & \multicolumn{1}{c}{\textbf{No. of Trainable Parameters}}  &   \multicolumn{1}{c}{\textbf{Training Time}} \\
    \midrule
    Single-task & 2.19 M & 10 ($\pm{2}$) min  \\
    Multi-task (Algo 1) & 2.20 M & 61 ($\pm{5}$) min\\
    Multi-task (Algo 2) & 2.20 M & 15 ($\pm{2}$) min
    \\\bottomrule
  \end{tabular}}
  \label{table: training_time}
\end{table}

\section{Discussion}

The results in Table \ref{table: single_vs_multitask} clearly confirms the impact of multi-task learning for the SI task with a relative improvement of 2.5\% over the single-task model. Over the two training algorithms, algorithm 2 has a slight edge in TV prediction. However, when training time for the two algorithms are considered (table \ref{table: training_time}), algorithm 2 has a considerable advantage by only taking nearly quarter of the time of algorithm 1. Hence for the subsequent experiments and comparisons we used the MTL model trained with algorithm 2. It should also be mentioned that in a previous work with developing a multi-corpus SI system \cite{seneviratne19_multicorpus}, a similar training procedure to algorithm 1 was used.

Figure \ref{fig:tv_plots} and figure \ref{fig:tv_plots_2} shows the ground-truth TVs and the predicted TVs from the multi-task and single-task models. The key difference between the two figures is that figure \ref{fig:tv_plots} shows the TVs which characterise the constriction degree of articulators, whereas figure \ref{fig:tv_plots_2} shows TVs which characterizes the constriction location. It is usually observed that SI systems tend to do better with constriction degree related TVs compared to ones which capture constriction location mainly due to the fact that the same speech sound can be produced with different vocal tract configurations (speaker-dependent characteristics). The same can be observed with the PPMC scores for each TV in Table \ref{table: single_vs_multitask}. But an interesting observation is that the multi-task models mostly improve in estimating location related TVs with respect to the single-task model suggesting that learning the phoneme mapping is helping the SI task with additional subject-dependent information.   

Table \ref{table: baseline_models} shows that the proposed MTL based SI system achieves the best PPMC scores over the existing SI systems on the HPRC dataset. The results also suggest that the BiGRNN and the CNN-BiLSTM models clearly outperform the conventional feed-forward neural network models in the SI task. Moreover, with the ablation study in section \ref{sec:ablation}, we show the importance of multi-task learning (i.e. learning a related, shared task) on improving the SI systems for TV prediction. This also suggests that with the joint loss $L_{joint}$ optimization, an additional hyper-parameter $\alpha$ needs to be fine-tuned properly to achieve the best results.

Finally, it should also be highlighted that the proposed MTL based SI system only uses phoneme transcriptions for training. At the time of inference, only the acoustic features are needed and it draws the key difference between the proposed SI system and the SI systems using both phoneme and acoustic features as inputs. 

\vspace*{-6pt}
\section{Future Work}

The lack of larger articulatory datasets for training DNN based models is a key challenge in developing generalizable SI systems. One of the envisions of developing a MTL based SI system lies with the idea of tapping into larger, existing datasets of audio, phonetic transcriptions (e.g. Librispeech \cite{librispeech}). The authors wish to work on transfer-learning and model adaptation paradigms to improve the current MTL framework by pre-training the models with existing corpora of audio, phonetic transcriptions.   

\vspace*{-5pt}
\section{Acknowledgements}

This work was supported by the National Science Foundation grant \#1764010

% This work was supported by Advanced ERC Grant NEUME 787836 and Air Force Office of Scientific Research and National Science Foundation grants

\bibliographystyle{IEEEtran}

\bibliography{mybib}

\end{document}